# Finding the proper node ranking method for complex networks

Senbin Yu[1], Liang Gao[1, 2, *], Yi-Fan Wang[1]

1. Institute of Transportation Systems Science and Engineering, MOE Key Laboratory of Urban Transportation System Theory and Technology, State Key Laboratory of Rail Traffic Control and Safety, and Center of Cooperative Innovation for Beijing Metropolitan Transportation, Beijing Jiaotong University, Beijing 100044, China
2. Yinchuan Municipal Bureau of Big Data Management and Service, Yinchuan 750011, P. R. China

**Abstract:** Ranking node importance is crucial in understanding network structure and function on complex networks. Degree, h-index and coreness are widely used, but which one is more proper to a network associated with a dynamical process, e.g. SIR spreading process, is still unclear. To fill this gap, $a$, which is extracted from the fitting function ($f(x) = 1 - \frac{1}{e^{2a(x-b)}+1}$) of the average number of nodes in each radius of the neighborhood of a node, is proposed. Experiment results which are carried out on twenty real-world networks show that $a$ can classify which of the three measures (degree, h-index and coreness) is more proper to a network in ranking node importance. We also find that $[b/3]$ is a good indicator for forecasting the optimal radius of the neighborhood of a node in ranking node importance for a given network. To the best of our knowledge, it is the first solution of this interesting and open issue. Furthermore, by extending the range of neighborhood where we construct an operator $\mathcal{H}$ on of a node, we propose a new method to quantify the importance of a node. The ranking accuracies of most networks can be improved when the radius is increased from 0 to its forecasting optimal radius and the improvement, for the best case, reaches up to 111%. The performances will reduce on half of the networks studied in this paper if we roughly extend the radius of the neighborhood. Our work deepens the understanding of how to find out the proper node ranking method for complex networks. The proposed methods bridge the gaps among network structure and node importance, and may have potential applications in controlling the outbreak of disease, designing of optimal information spreading strategies.

## 1. Introduction

Complex networks characterize the nature of interactions in a real-world system including social, economic, technological networks and so on. Indeed different real networks exhibit heterogeneous characteristics with nodes play various roles in structure and function [1]. Identifying the critical nodes associated with the specific dynamics gains attention of a growing number of researchers from different disciplines [2-5], since it's the first step to better control the outbreak of epidemics [6, 7], conduct successful advertisements for e-commercial products [8], prevent catastrophic outages in power grid or the Internet [9], optimize the use of limited resources to facilitate information propagation [10], discover drug target candidates and essential proteins [11], and design strategies for communication breakdowns in human and telecommunication networks [12].

Numerous researchers focus on how to rank node importance from epidemic dynamics [13-17]. Degree, as the simplest and intuitive indicator, believed that the number of a node's linkages can measure the importance of a node [14, 18]. While, Kitsak et al. [13] argued that if a highly connected node exists at the periphery of a network, it will have a minimal impact in the spreading process through the core of the network, whereas a less connected person who is strategically placed in the core of the network will have a significant effect that leads to dissemination through a large fraction of the population [13]. So the position of the node identified by k-core decomposition analysis [19] plays a more critical role in epidemic spreading. A larger coreness value indicates that the node is more centrally located in the network. As a good index for nodes' influence, coreness has been applied in many real networks [13, 20, 21] and improved, as a benchmark metric, to promote the ranking accuracy and reliability [15, 17, 21-30].

Recently Lü et al discuss the extension of the h-index concept, which was originally used to measure the citation impact of a scholar or a journal, to quantify how important a node is to a network [31, 32]. They construct an operator $\mathcal{H}$ on a group of reals that returns a node's h-index when acting on its neighbors' degrees. The h-index of a node is defined to be the maximum value $h$ such that there exist at least $h$ neighbors of degree no less than $h$ [16]. They find that the h-index outperforms both degree and coreness in several cases when measures a node's influence.

Degree, h-index, and coreness are seemed to be independent but actually interrelated. The iterative k-core decomposition process requires global degree information of the network. In contrast, the degree is a simple local index. As a tradeoff, by applying the $\mathcal{H}$ operator to each node, the returned value soon converges to coreness, that is, in terms of this operator, degree, h-index, and coreness are its initial state, intermediate state and steady state [16]. The degree can be defined as the zero-order h-index ($h^0$) and coreness is $\lim_{n\to\infty} h^n$, where n is the n-order h-index. And at the same time, they owns widely applications for identifying influential nodes in complex networks. They were always considered as benchmark centralities compared with their own improved methods [13, 15, 17, 21, 25, 33-35] or some new methods from the other perspectives [36-39].

But one seldom observes two phenomena when identifies and ranks influential nodes by them: The relative size of their performance are stability using various infection probability under numerical analyses of the susceptible-infected-removes spreading dynamics (SIR) when the infected rate $\beta$ is larger than $\beta_c$; None of them can always offer the most exact ranking than the other two methods in all the networks [16, 40, 41]. Each method, for example, h-index, has its advantage to some certain type of networks. We test six different types of networks including twenty real-networks in different infection probability (see Supplementary Table 1 and Table 2 for detail information) which are consistent with the above-mentioned characteristics. So according to the best performance which can be given by one method among degree, h-index and coreness associated with SIR model, a network may be a degree advantage-type network (DAN) or h-index advantage-type network (HAN) or coreness advantage-type network (CAN).

The influence of a node is largely affected and reflected by the topological of the network it belongs to. In fact, the majority of known methods in identifying influential nodes only make use of the structural information, which allows the wide applications independent to the specific dynamical process under consideration [1]. As we know, degree, h-index, and coreness can be obtained by only considering the degree information in different scale of the neighborhood. Motivated by it, we suppose that the characteristics in the distribution of the average number of nodes in each $r$-step neighborhood of a node (ANRN) can classify the advantage-type networks. We find a parameter $a$ extracted from the fitting function of ANRN (FANRN) is proportional to the maximum variation rate in FANRN can classify the DANs, HANs and CANs compared with seven classic properties of network structure.

Moreover, the other parameter $b$ of FANRN is a good indicator ($[b/3]$) for forecasting the optimal radius ($r^{opt}$) of the neighborhood information which is utilized for a node in ranking node importance. To the best of our knowledge, it is the first solution of this interesting and open issue. Furthermore, we extend node $i$'s range of neighborhood where we construct an operator $\mathcal{H}$ on and introduce a $mh_i^r$-index to describe node $i$'s importance. Interestingly, by applying forecasting $r^{opt}$ ($r^{f-opt}$), the performance of $mh^{r^{f-opt}}$-index for identifying the influential nodes can be improved compared with $mh^0$-index (degree) and $mh^1$-index (degree) on most real-networks in this paper. In contract, the ranking accuracy will decline in half of the networks if $r$ is set as an arbitrary value (for example $r = 2$).

## 2. Results

### 2.1 The best method for identifying influential nodes

Epidemic spreading is one the most significant behaviors in complex networks which can be used to characterize most common process in many domains such as the spreading of infectious disease [42], the diffusion of microfinance [43] and the propagation of traffic [44]. In order to evaluate the best method for a network by spreading influence, we employ the SIR model to simulate the spreading process, where the influence of nodes are denoted by spreading vector $R$, computed by the average number of recovered and infected nodes at the steady stated of SIR process after 200 independent simulations, and each simulation begins with only one node as the single infection seed (see Methods for single seed SIR model in details). We apply Kendall $\tau$ ($\tau_b$) correlation coefficient [45] to quantify prediction accuracy, where this measurement can well abstract the correlation. $\tau_b$ lies in $[-1, 1]$, the greater absolute value of $\tau_b$ implies a higher correlation between two sample vectors (see more on Kendall $\tau_b$ in Methods). Higher correlation between the method score vector and the spread range vector indicates better prediction accuracy. To find the best method among degree, coreness and h-index for a network, we calculate $\tau_b^d(R_d, R)$, $\tau_b^c(R_c, R)$ and $\tau_b^h(R_h, R)$ respectively. A network whose ranking accuracy obtain the best performance by h-index is a HAN when the spreading rate $\beta$ is set as the same value. One can get a DAN and a CAN in the same way.

In this work, twenty real-world networks (including six types) are considered whose node size range from 34 to 58228. The networks types studied are infrastructure networks (US Air [46] and Power Grid [47]), Internet networks (Oregon1-010505 [48], p2p-Gnutella06 [49, 50], p2p-Gnutella08 [49, 50], p2p-Gnutella09 [49, 50]), Neural network (C.elegans [51]), communication network (Email e, Terrorist [52]), social network (loc-Bright [53], Karate club [54], Dolphins [55], PGP [56] and Hamster [57]) and collaboration network (ca-AstroPh [49], ca-Condmat [49], ca-GrQc [49], ca-HepPh

[49], ca-HepTh [49], NetSci [58]).

The topological of a network can affect and reflect the nodes' influences. In ref [37], the authors find that a node with more nodes in its 2-step neighborhood which can be reached out to from 1-step neighbor does infect many neighbors after infected this node by its connecting edges, while the spread range will cease quickly if its 2-step neighbors are seldom. Ref [17] developed a new general framework to rank nodes through gathering neighbor's information combined with a priori knowledge iteratively. And a local h-index centrality based on the node itself and its neighbors are proposed to identifying influential nodes in complex networks [40].

Owing to calculate the degree information using a different radius of the neighborhood, we can get the score $k_i$, $ks_i$ and $h^i$ of node $i$ by degree, coreness and h-index respectively. So on the base of previous studies, we research the characteristics in the distribution of the average number of nodes in each n-step neighborhood of a node (ANRN) which we suppose can classify the advantage-type networks using different methods.

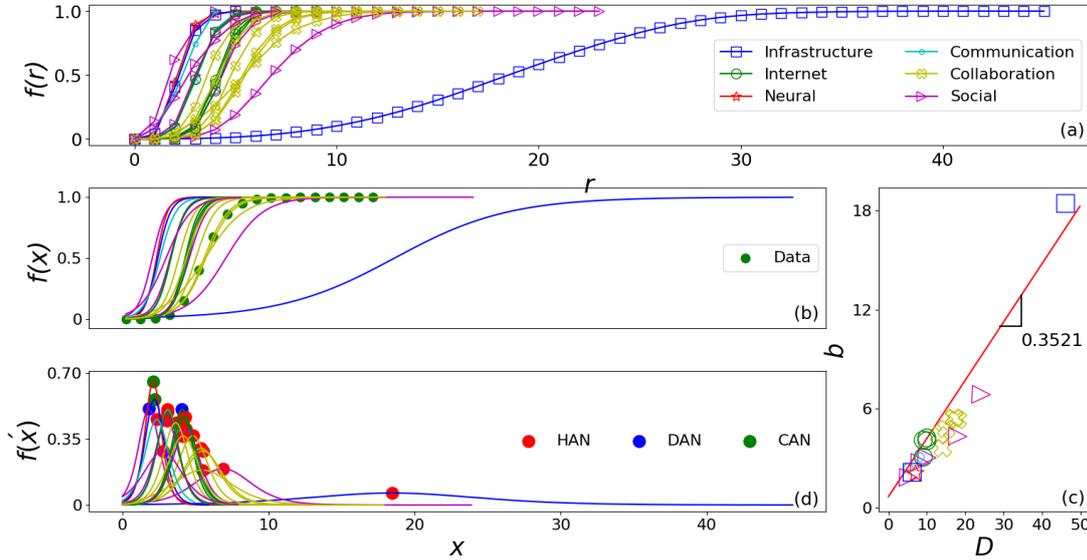

Figure. 1. The distribution of ANRN and its fitting function. (a) The distribution of ANRN on six types of real-world networks. All the twenty networks have a similar distribution shape of ANRN. (b) A unified function $f(x) = 1 - \frac{1}{e^{2a(x-b)}+1}$ can effectively fit the distribution of ANRN by least squares curve fitting method. The green point is the real data from an Internet network whose matching is better intuitively. (c) The parameter b and the diameters of six type networks display linearity relationship which has a slope of about 0.3521. (d) The graph of the derivative of FANRN (DFANRN). The red round, blue round and green round are all on the crest of curves which are HANs, DANs or CANs, respectively. The infected probability is $1.5\beta_c$.

As shown in Fig. 1a, we can find that the ANRN of different types of networks can fit well by the function $f(x) = 1 - \frac{1}{e^{2a(x-b)}+1}$ in Fig. 1b, where $x$ is a unique variable, $a$ and $b$ are two parameters in FANRN (see more detail about the goodness of fit see Supplementary III Table 3). The shape of this function has rotational symmetry with respect to the point $(b, 0.5)$ which is the most important vertex in the function. The change rate of function reaches the maximum $a/2$ is always proportional to the value of parameter $a$ at the symmetric point. Then the derivative will decrease when $x$ is larger than $b$. Interesting, as shown in Fig. 1 (d), we observe that the range of $a$ which a network belongs to is highly relevant with the network's best method (one method among degree, coreness and h-index) for ranking nodes' importance. The networks whose $0 < a < 1$ can be includes HANs. CANs belong to the range $a > 1.1$ and the rest scope of $a$ ($1 \leq a \leq 1.1$) is employed for DANs. To see whether $a$ can classify the different method advantage-type networks, we compare it with seven representative network properties are examined in this paper. The properties of the real-world networks are calculated including average degree ($\langle k \rangle$), maximum degree ($k_{max}$), maximum k-shell index ($KS_{max}$), degree assortativity ($\varphi$), clustering coefficient ($C$), density of a graph ($d$), and diameter of a graph ($D$) (see more dataset description in Supplementary Section I Table 1). For better observation, we describe the different values of $a$ though different network types displaying on the Fig. 2. The seven classic network properties cannot divide the different advantage-type networks. DANs, HANs and CANs mix together in a different range of network property whose best method are same for identifying influential nodes. But you can easily observe three layers in Fig. 1 (h) and in Supplementary Section IV Fig. 6 and Fig. 7 when using different infected rates. So the value of $a$ indicates the best method which can help us find the proper method for identifying the influential node.

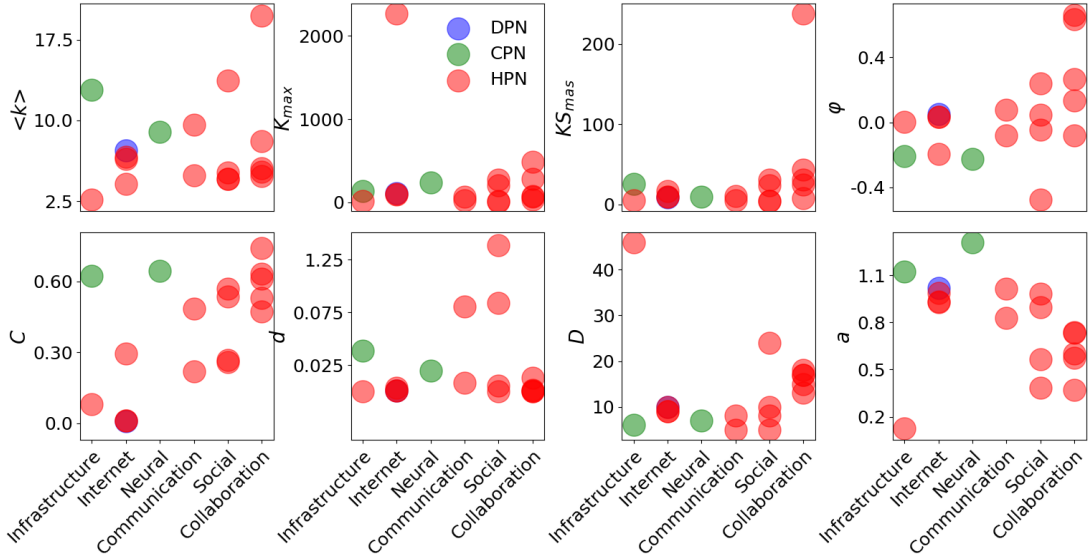

Figure. 2. Evaluation of the relationship between network property and advantage-type method. The spreading rate $\beta$ is set as $\beta = 1.5 \beta_c$. In many cases h-index outperforms both degree and coreness [16], so most networks are HANs. Few networks are DANs or CANs owing to their own shortcomings [1, 13].

## 2.2 The optimal radius

Increasingly evidence shows that propagation capability largely depends on information from neighbors. As for macroscopic information, the spreadability is hard to be expressed through single node information, but it can be captured better by the neighbor based centralities [15, 16, 33, 40, 59, 60]. An ideal centrality is better to contain more neighbors' information and reflect more global structural features of the target network [17]. Many centralities based on neighbor's information have been proposed, such as semi-local [59], extended neighbors' coreness [15], leveraging local h-index [40], improved neighbors' k-core [60], neighborhood centrality (NC) [33], h-index [16], iterative neighbor information gathering (ING) [17] and the number of possible infection paths in the neighborhood [61] providing us some accurate and reliable ranking results. Intuitively, the larger neighborhood is taken into consideration, the more accuracy we can predict the spreading outcome of a node.

But first, it is a challenging task to collect the complete network information in some real-world networks [20], due to the large amount and the temporal and spatial change of the data, such as Twitter and Facebook. Analyzing a large neighborhood seems unfeasible in such networks. Second, there is a significant relationship in behaviors between a node and its direct neighbors, and up to the 3-step neighbors [62]. In the case of the large size of neighbor, when the scale can include a large fraction of the population, the role of individual nodes is no longer important, independently of whose influence is more important. However, sometimes a small size of neighbors have inadequate information to represent a node. And Liu argued that considering the 2-step neighborhood of nodes is a good choice that balances the cost and performance, which is not the case of the larger the better [33].

In ref [33] degree and coreness were taken into consideration as the benchmark centralities to study the performance of NC. A saturated effect was discovered when considering the neighbors within 4-step. The NC outperforms the benchmark centrality in ranking node influence. Though the mentioned NC has a certain extent researched about the saturation effect, it only guesses that this effect may relate to the spreading radius of nodes and there is not a precise value of optimal neighbor size for each network whose optimal neighborhood may not be 2-step.

So which size the neighborhood is can best describe a node's influence? Do the optimal sizes are different in different kinds of the network? Can the network structure reflect the optimal size?

Without any doubt, the optimal size of a neighborhood depends on three factors: benchmark centrality, network topology, and dynamics process obviously [17, 33]. In order to investigate the optimal neighborhood of node $i$ in a network, we define a simple method by the accumulation of different layers of neighbor information (AN) in its $r$-step neighborhood shown as: $AN_i^r(\mu) = \sum_{j \epsilon \Gamma_i^r} \mu_j$, where $\mu$ is the benchmark centrality, $r$ is the shortest path length from node $i$, $\Gamma_i^r$ is the set of 0-to-$r$ step neighborhood of node $i$. Here we utilize degree, coreness and h-index as the benchmark centrality to study the performance of AN by different $r$ (In case there exists more than one connected component in a network, $D$ is calculated by the giant connected component in the network). So $AN^r$ encodes its $r$-step neighborhood in its definition.

In Fig. 3, we discuss the impact of $r$ on the performance of $AN^r(\mu)$ by calculating the Kendall $\tau_b$ between method score and spreading influence by SIR model. The benchmark measure is one of degree, coreness and h-index. We can see that in general, the best performance of identifying influential nodes by range $r$ is single in each network. In US Air, C.elegans, Email, Terrorist, Dolphins, Hamster, ca-HepPh, $AN^1(\mu)$ can off the best ranking accuracy. In ca-GrQc, ca-HepTh and Netsci the largest $\tau_b$ lies at $r = 2$. In Power Grid, the largest $\tau_b$ is achieved within the six-step neighborhood. And $AN^r(\mu)$ almost has a same characteristic of ranking correlation where the optimal $r$ ($r^{opt}$) lies on and the trend of the change of the $AN(\mu)$ counted by different benchmark method is very similar on most networks.

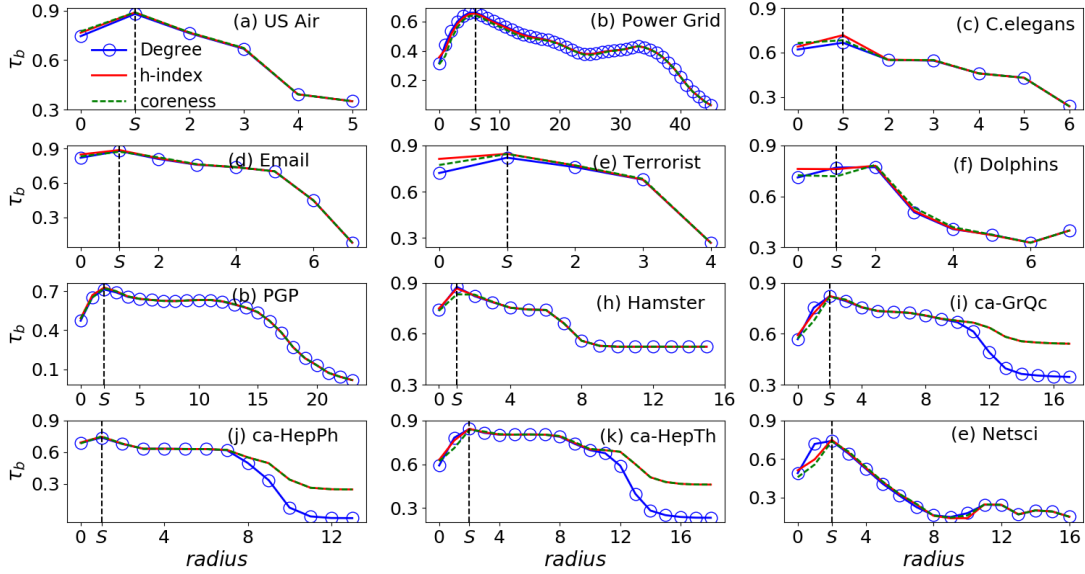

Figure. 3. The Kendall $\tau_b$ between $AN^r$ and the node influence index $R$ for undirected networks. The value of $r$ ranges from 0 to $D$. The infected probability $\beta$ is $2.5\beta_c$. The blue circles, red lines and green dash line represent degree, h-index and coreness, respectively. The black dash lines emphasize the forecasting optimal radius counted by $[b/3]$. In addition, the influence $R$ of a node is quantified using the average number of removed nodes after the dynamics over 200 independent runs.

As we known, the FANRN has rotational symmetry with respect to the point $(b, 0.5)$. The change rate of functions reach the maximum $a/2$ and then decrease with the adding of $x$ which actually represents the radius. It means the rate of increasing neighbor information decreases. Based on the analysis about the optimal $r$ of the neighborhood, we conjecture the value of $b$ has close contact with the optimal $r$ of nodes for quantifying the nodes' importance.

We find that the $r^{opt}$ can be counted by the round method to round up and round down $[b/3]$ ($s$). As shown in Fig. 3, $s$ obtained by each network's structure information can predict the $r^{opt}$ accurately in a different type of networks on different benchmark methods. For the exception Dolphins, $AN^0(h\text{-index})$ performs the best, while $AN^s(h\text{-index})$ is still a suboptimal solution. We further compare the relationship between $r^{opt}$ and $s$ on more networks when the infection probability is set as different values. As shown in Supplementary Section V from Fig. 9 to Fig. 13, $s$ is a good indicator of $r^{opt}$ too.

Most networks' optimal layers in this paper are greater than 1 and it is a relatively stable value in a wide range of infection probability for most networks (see Supplementary Section V Fig. 8 for detail). That's why some previous work can obtain good performance after adding the information of direct neighbors [15, 33, 40]. The performances of different methods first increase to largest values and then decrease in almost all the network along with radius. This conclusions also explain the correctness of a basic assumption in a way — a spreader with more connections to more influential nodes will also be influential in many researches [15, 26, 33, 34, 36, 37, 40, 41, 63, 64] who take count of first-step neighbors' benchmark centrality for identifying the influential nodes. The $r^{opt}$ is always equal or bigger than one, so a method who make a summation of the node's and its direct neighbors' information can always outperform than a method just includes each node itself.

As mentioned earlier, the benchmark centrality, network topology and dynamics process [17, 33] determine the $r^{opt}$ of a network. As we take the benchmark centrality (degree, coreness or h-index) into account, we find that counter-intuitively, it is not a significant difference among them. Hence we believe that the network topology plays a leading role in this problem when we use relatively big range values for $\beta$.

Furthermore, the diameter $D$ has a great influence about $r^{opt}$ and there exists a linear correlation between them counted by different real-world networks. Thus we get a simple estimate on the $s$ for its applications. That is $s^r \cong [(0.3521D + 0.8175)/3]$ where $s^r$ is the roughly estimating value, and $D$ is a single parameter which can be calculated by some classic algorithms like the Floyd–Warshall algorithm [65] and Williams algorithm [66] in low time complexities. The $s^r$ of each network remains much as $s$ in Supplementary Section II Table 2. So it is beneficial for the application of $s$ for identifying influential nodes.

**2.3 An simple application**

In this section, we try to make an application on our conclusions to find a better method for identifying the influential nodes.

As we know, h-index of a node is defined as the maximum integer $h$ such that the considered node has at least $h$ neighbors whose degree are greater than $h$ [16]. Higher h-index indicates that the node has a number of neighbors with high degree.

In ref [37], researches find that a node with more nodes in its 2-step neighbors which can be reached out to from 1-step neighbors does infect many neighbors after infected this node by its connecting edges, while the spread range will cease quickly if its 2-step neighbors are seldom. Ref [17] developed a new general framework to rank nodes through gathering neighbor's information combined with a priori knowledge iteratively. And a local h-index centrality based on the node itself and its neighbors are proposed to identifying influential nodes in complex networks [40].

Actually, by combing the knowledge from the results describe above, we try to extend the range of neighborhood where we construct an operation $\mathcal{H}$ on to see whether the new method can improve the performance for quantifying the nodes' influence. Here the direct neighbors can be extended to multilayer neighbors. And a $mh^r$-index of node $i$ is defined to be the maximum value $h$ such that there exists at least $h$ nodes of degree no less than $h$ in $i$'s multilayer neighbors. We define the multilayer neighbors as a $r$-step neighborhood like $AN^r$. So for node $i$ of an undirected simple network, we can get a new sequence $mh_i^0, mh_i^1, mh_i^2, \cdots$ by the different $r$-step neighborhood. We define $mh_i^0 = k_i$ to be the degree of node $i$ and obviously $mh_i^1$ is the h-index of node $i$ indeed.

Given the radius $r$ of the neighborhood ($0 \leq r \leq D$), we have two sequences associated with the $|V|$ nodes: the $mh^r$-index $mh_1^r, mh_2^r, \cdots, mh_{|V|}^r$ and the influence $R_1, R_2, \cdots, R_{|V|}$. To quantify to what extent the $mh^r$-index resembles node influence values, we also apply the Kendall $\tau_b$ which means a stronger correlation between the two sequences (see methods for details). So the $r$ is an influence factor for the performance of identifying importance nodes.

We apply $s$ calculated by each network structure to $mh^r$-index and compare $mh^s$-index with degree or h-index for testing performance. As shown in Fig. 4 the improvement is a positive number in describing the node influence calculated by the difference of Kendall $\tau_b$ has been made in the most networks. In some case (Power grid, PGP and ca-HepTh), the performance measured by h-index can increase by more than 50% if we know $s$ of a network. p2p-Gnutella06 and Karate club are two exceptions whose ranking accuracy has slightly decreased (less than 5%). Note that both of them belong to DANs whose performance is better than h-index for ranking the influential nodes when the infected rate is $1.5\beta_c$ and decrease with the incensement of $r$. Relatively all the CANs and HANs' performances can be promoted in various degrees.

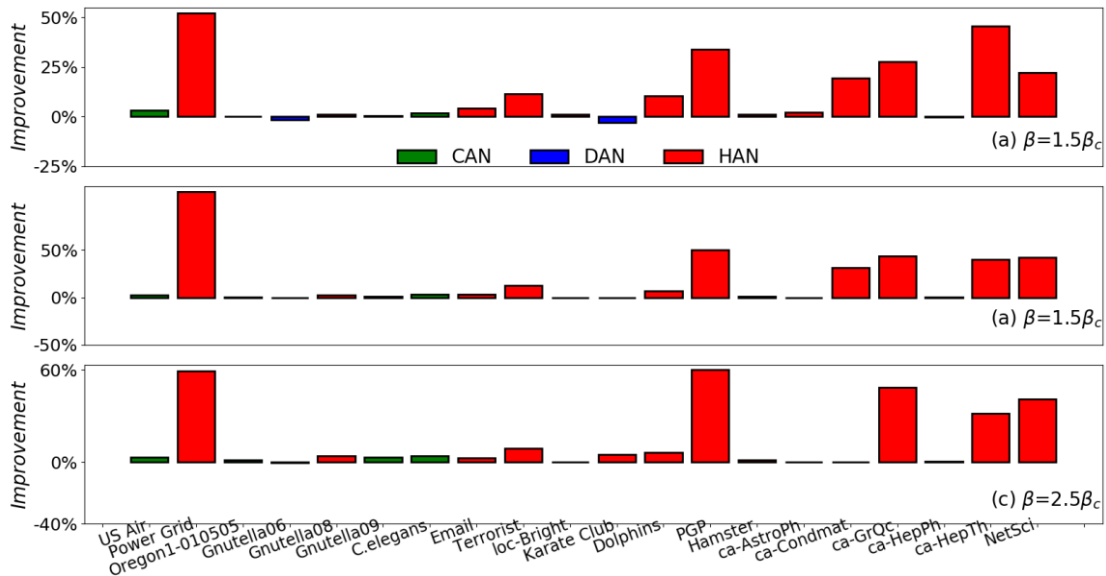

Figure. 4. The percentage of performance changes compared by $\tau_b(R_{mh^o}, R_{SIR})$ and $\tau_b(R_{mh^s}, R_{SIR})$. Here, we make $mh^0 = degree$ to be the zero-order of $mh^r$. The green bars, red bars and blue bars represent the case of CANs, HANs and DAHs, respectively.

But if one has no knowledge about $s$ of a network and arbitrarily extends the $r$ (for example $r = 2$) when contracts an operation $\mathcal{H}$, the property for identifying influential nodes may greatly reduce in all DANs, CANs and part of HANs as shown in Figure 5.

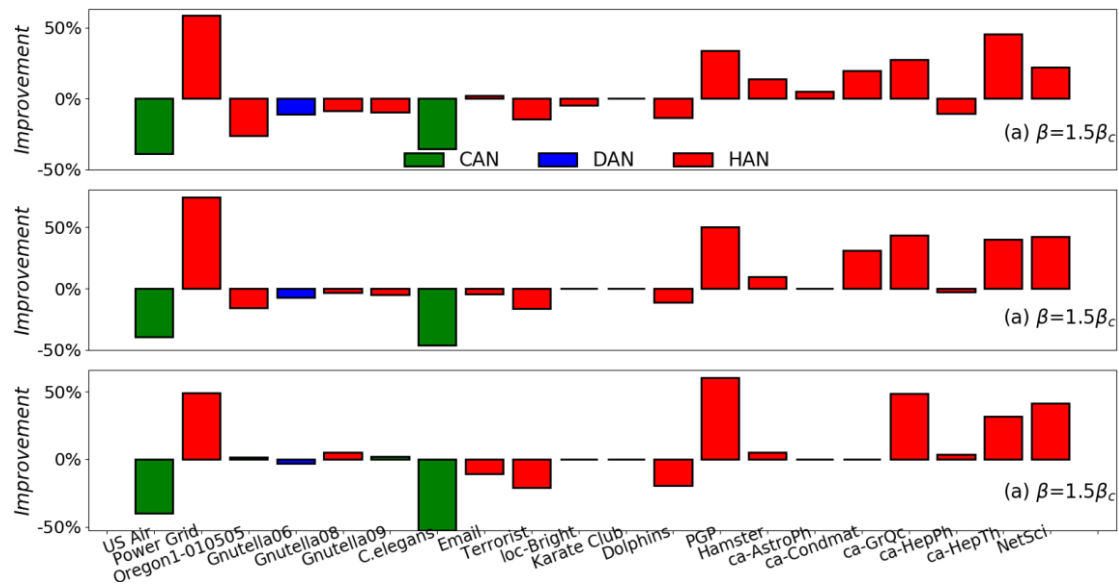

Figure. 5. The percentage of performance changes compared by $\tau_b(R_{mh^0}, R_{SIR})$ and $\tau_b(R_{mh^2}, R_{SIR})$. Here, we $mh^0 = $ degree to be the zero-order of $mh^r$. The green bars, red bars and blue bars represent the case of CANs, HANs and DAHs, respectively.

Similarly, we further compare the performance changes by $r$ from 1 to some other value and obtain the same conclusions as shown in Supplementary Section VI Fig. 14 and Fig. 15. And So we believe that $s$ can be a better indicator to finding the $r^{opt}$ when identify the influential nodes using neighborhood information which associated with SIR model as shown in Supplementary Section VI figure 16 and figure 17. But there is still a shortcoming for the application of $s$ which is not very convenient to figure out. As mentioned in section II, we offer $s^r$ of a network by a simple way. By applying the $s^r$ for quantifying the nodes' importance in $mh^r$-index, the performance can also be improved in most networks in Supplementary Section VI Fig. 18 which show $s$'s great potential for the application of identifying influential nodes.

## 3. Discussion and conclusions

Node importance or influential node identification for complex networks is still an open issue. From the viewpoint of machine learning, this task is a kind of unsupervised learning, i.e. learning process without a guide [17]. Identifying and ranking influential nodes in the dynamical process play a crucial role in understanding network structure and function. Degree, h-index, and coreness are widely used measures and there exists a close relationship among them [16]. Numerical analyses of the susceptible-infected-recovered spreading dynamics on real networks suggest that the h-index is a good tradeoff that in many cases can better quantify node influence than either degree or coreness. This does not indicate that h-index has an absolute advantage over other measures. Actually, in some cases, h-index is worse than the other two measures.

We find that each network in this paper has its own advantageous method to quantify its node influence. For example, although we know that degree is not an accurate centrality measure in quantifying node influence, it is still a useful estimation of node importance in p2p-Gnutella06 in case of different $\beta$ on SIR model. So in order to offer more exact rankings, we need a basic judgment to identify which one is more proper to a network and then do some further promotions and applications.

Based on the discussion about how to calculate degree, coreness, and h-index, we suppose the distribution of the average number of nodes in each $r$-step neighborhood of a node (ANRN) can classify the advantage-type networks. We find that the ANRN for twenty networks in this paper can be better fitted by an extended tanh-function ($f(x) = 1 - \frac{1}{e^{2a(x-b)}+1}$), where $a, b$ are parameters and $x$ ($x > 0$) is the variable in this function. And the value of parameter $a$ can better classify the advantage-type network of the method. By the way, although Newman had solved the calculation of a number of neighbors based on generating functions, but his result depend on the detailed structure of the graph and assumed that all vertices are reachable from a randomly chosen starting vertex [67]. In general, however, this will not be true. It is not suitable for us because of such shortcomings and we will try to give an explicit and unified formulas in our further work.

Then we discuss the optimal $r$ of the neighborhood when we utilize the neighborhood information to identify the node's importance. By counting the neighbor information in the node's $r$-step neighborhood, we find that $[b/3]$ ($s$) can forecast the $r^{opt}$ regardless of the benchmark method (degree, h-index and coreness) in different infected probability $\beta$.

$s$ is also a useful indictor to predict the scale of the neighborhood in a new method $mh^r$ we proposed. If the $s$ has been used, we can promote the performance of ranking nodes' importance on almost all the networks in this paper. In contract, if one arbitrarily determines the range of neighborhood (for example $r = 2$) the accuracy of node importance will decrease in over half of networks in this paper. In fact, the $s$ is also the same as the $r^{opt}$ by ranging $r$ on $mh^r$ as shown in Supplementary Section VI Fig. 16 and Fig. 17.

Actually, the $s$ of a network obtained from the fitting function is not easy to calculate for a large scale network. For simplicity, an approximate computational method for the optimal $r$ is proposed which need only the diameter ($D$) of a network and defined as $s^r = [(0.3521D + 0.8175)/3]$. To test the performance, we compare the $mh^{s^r}$-index with $h^0$-index (degree) as summarized in Supplementary VI Figure 18, the result suggest that the $mh^{s^r}$-index are still very competitive.

Because coreness cannot be used in some classical modeled network, such as Barabási–Albert (BA) networks [68] and tree-like networks, where the coreness values of all nodes are the very small and indistinguishable [1]. We mainly consider six type real-world networks (including twenty networks) without taking the modeled networks into account here in this paper.

The proposed methods bridge the gaps among network structure and node importance and may have potential applications in finding an exact expression for the average neighbors of a vertex, controlling the outbreak of disease, designing of optimal information spreading strategies.

## 4. Methods

Given a network $G = (V, E)$, $N = |V|$ is the number of nodes, and $M = |E|$ is the number of edges. Let $e_{ij}$ represent the edge connecting node $i$ and node $j$, and $\Gamma_i$ denotes the set of neighbor nodes of node $i$. The degree $k_i$ is the number of links node $i$ carries.

**Degree, coreness, h-index**

The ***degree*** is the simplest indicator to quantify node importance. It focuses on a number of links per node and believes that the most connected nodes are hubs.

The ***coreness*** is obtained in the K-shell decomposition process. Each node will be assigned to a K-shell index by the process recursively pruning nodes with degree less than or equal to $k$. The pruning process continues until all nodes in the network are removed. As a result, each node is associated with one K-shell index.

The ***h-index*** of node $i$ in a network is defined as the maximum value $h$ such that there are at least $h$ neighbors of degree larger than or equal to $h$. It is an operator acts on a finite number of integer $(k_{j1}, k_{j2}, \cdots, k_{jk_i})$ and returns an H-index value of node $i$, where $k_{j1}, k_{j2}, \cdots, k_{jk_i}$ are the degrees of node $i$'s neighbors. Hence, we set $H_i$ as the H-index of node $i$ as follow:

$$H_i = \mathcal{H}\{k_j | j \epsilon \Gamma_i\}, \tag{4}$$

where $k_j$ is the degree of neighbor node $i$.

**Single seed SIR model**

The single seed SIR model is applied to study the spreading process. In the single seed SIR model, all nodes are initially in the susceptible state (S) except for the only one seed node in the infectious state (I). At each time step, the I state nodes infect their S neighbors with probability $\beta$ and then enter the recovered state (R), where they become immunized and cannot be infected again. This process is repeated until there are no infected nodes in the network. The number of recovered nodes gives the final infected scope of the seed node, which is adopted to measure the spreading influence of the seed node. The higher $\beta$, the larger population infected, no matter where it originates. And the critical infection probability $\beta_c \sim \langle k \rangle / \langle k^2 \rangle$ where begins the infectious disease outbreak will be used in this paper.

**Evaluations**

*Kendall correlation coefficient* **[45]** is adapted to measure the consistency between two rankings. Given $R_\mu$, the rank vector of measure $\mu$, and $R_{SIR}$, that of the single seed SIR model, the Kendall $\tau_b$ correlation coefficient is defined as

$$\tau_b(R_\mu, R_{SIR}) = \frac{n_c - n_d}{\sqrt{(n_0 - n_1)(n_0 - n_2)}}, \tag{8}$$

where $n_c$ is the number of concordant pairs, $n_d$ is the number of discordant pairs. And $n_0 = n(n-1)/2$, $n_1 = \sum_i t_i(t_i - 1)/2$, $n_2 = \sum_j u_j(u_j - 1)/2$, where $n$ is the size of rank vectors and $t_i$ and $u_j$ are the number of tied values in the $i_{th}$ and $j_{th}$ group of ties, respectively. Since all measures are evaluated by $R_{SIR}$, $\tau_b(R_\mu, R_{SIR})$ will be denoted by $\tau_b(\mu)$ for short.

Let us suppose that set $S^c$ measured by a characteristic of each network's structure in ascending order and the corresponding $S^p$ recoding the best method of each network are all known.

**Acknowledgments**

The authors thank for support from the National Natural Science Foundation of China (No.71571017, No.91646124, and No.71621001), and support from the Fundamental Research Funds for the Central Universities (2015JBM058).